\documentclass[12pt]{llncs}
\usepackage[letterpaper,top=1in,bottom=1in,left=1in,right=1in]{geometry}
\usepackage{aeguill,xspace} \usepackage[T1]{fontenc}
\usepackage[latin1]{inputenc} \usepackage{graphicx}
\def\keywords#1{\begin{quote}\small\textbf{\keywordname} #1\end{quote}}
\newlength{\figheight} 

\let\includefigurea\includefigure \let\includefigureb\includefigure
\def\includefigurebig#1#2#3{\begin{figure}[tp]\centerline{\includegraphics[height=1.5\figheight]{#1}}\caption{#2}\label{fig:#3}\end{figure}}

\def\up#1{\raisebox{0.8ex}{\mbox{\scriptsize{#1}}}}
\newcommand{\jxta}{\textsc{JXTA}\xspace}

\title{Performance Analysis of Publish/Subscribe Systems}
\author{Heithem Abbes\inst{1} \and Christophe C\'erin\inst{2} \and
  Jean-Christophe Dubacq\inst{2} \and Mohamed Jemni\inst{1}}
\institute{\'Ecole Sup\'erieure des Sciences et Techniques de Tunis,
  Unit\'e de recherche UTIC\\5, Av. Taha Hussein, B.P. 56, Bab Mnara,
  Tunis, \textsc{Tunisia} \\Tel: (+216) 71 496 066 \qquad Fax: (+216) 71
  391 166\\\email{heithem.abbes@esstt.rnu.tn, mohamed.jemni@fst.rnu.tn}
  \and
  LIPN --- UMR CNRS 7030 --- Institut Galil\'ee --- Universit\'e Paris-Nord\\
  99, avenue Jean-Baptiste Cl\'ement, 93430 Villetaneuse, \textsc{France}\\
  Tel: +33(0)1 49 40 35 78 \qquad Fax: +33(0)1 48 26 07
  12\\\email{\{christophe.cerin,jean-christophe.dubacq\}@lipn.univ-paris13.fr}}
\begin{document}
\maketitle
\bibliographystyle{plain}
\begin{abstract}
  The Desktop Grid offers solutions to overcome several challenges and
  to answer increasingly needs of scientific computing. Its technology
  consists mainly in exploiting resources, geographically dispersed, to
  treat complex applications needing big power of calculation and/or
  important storage capacity. However, as resources number increases,
  the need for scalability, self-organisation, dynamic reconfigurations,
  decentralisation and performance becomes more and more
  essential. Since such properties are exhibited by P2P systems, the
  convergence of grid computing and P2P computing seems natural. In this
  context, this paper evaluates the scalability and performance of P2P
  tools for discovering and registering services. Three protocols are
  used for this purpose: Bonjour, Avahi and Free-Pastry. We have studied
  the behaviour of theses protocols related to two criteria: the elapsed
  time for registrations services and the needed time to discover new
  services.  Our aim is to analyse these results in order to choose the
  best protocol we can use in order to create a decentralised middleware
  for desktop grid.
\end{abstract}
\section{Introduction}
The exploitation of new instruments, such as, very high energy
accelerators, telescopes and satellites in astrophysics, big data bases
of imagery in biology and medicine, numerous sensors in geology,
generated an important expansion of the needs in scientific
computation. Thus, it becomes necessary to establish new computing
infrastructures. On the other hand, the computer networks equipments
knew these last years, an important development of transmission speed
performances and devices became equipped by powerful processors and
important storage capacities.

These factors advantaged the emergence of new infrastructures, such as
grid computing, to respond to computation needs with an economic
cost. One variant of grid computing is the Desktop Grid where nodes are
merely desktop PCs. This category constitutes the setting of our work.

Research in grid has developed some specific software (middleware) for
the management of data and resources. The most Desktop Grid middlewares
are centralised. In this setting, our work consists in conceiving a
decentralised grid computing middleware based on peer-to-peer
systems. To realise this, we would profit from existing decentralised
peer-to-peer systems.

The service discovery in the Grid is among the principle challenges. For
instance, Globus middleware implemented the service publish/discovery
mechanism based on Monitoring and Discovery of Services
(MDS-2)~\cite{fknt02,fk99} which uses centralised register
server. Although MDS-2 solved the scalability problem using hierarchical
architecture, it is still vulnerable to single point of
failure. Moreover, adaptation to the dynamic feature of servers is
another challenge for MDS-2. Furthermore, another alternative consists
of using decentralised approach for service
discovery~\cite{cdkr02,xwy05}. Recently, P2P communities have developed
a number of fully decentralised protocols, such as
Bonjour~\cite{book-zeroconf,url-bonjour,url-zeroconf},
Avahi~\cite{url-avahi} and Pastry~\cite{rd01,url-freepastry} for
registering, routing and discovering in P2P networks. The core idea
behind these protocols is to build self-organised overlay networks when
nodes join the grid. On the other hand, it is important to know the
performance and the limits of such systems. In this context, several
experiments have been done in this work to analyse the performance of
Bonjour, Avahi and Free-Pastry.

We choose Bonjour and Avahi (two popular middlewares running on a local
area network) because our working context is the connectivity issues
that we are faced to when we try to share resources belonging to
different institutions. In this paper, we assume that we have a high
level middleware able to virtualise the network (we have no more
problems with firewalls and NAT) and we are able to run Bonjour and
Avahi on top of such middleware. Instant Grid~/ Private Virtual Cluster
(see~\cite{rmnc06,url-instantgrid}) is one of the candidate for network
virtualisation. Its main requirements are: 1) simple network
configuration 2) no degradation of resource security 3) no need to
re-implement existing distributed applications. Under these assumptions,
it is reasonable to check if Bonjour and Avahi can scale up.

This paper is organised as follows. In section~\ref{section:grid}, we
remind the notion of large scale distributed systems by focusing on grid
computing and peer-to-peer systems. Then, in the
section~\ref{section:desktop-grid}, we illustrate the notion of desktop
grid. In section~\ref{section:peer-to-peer}, we highlight the advantage
of peer to peer systems in building a new decentralised middleware for
grid computing. In section~\ref{section:experimental-setup}, we describe
the experimental setup used to evaluate the performance of Bonjour,
Avahi and Free-Pastry. In section~\ref{section:registration-analysis}
and section~\ref{section:discovery-analysis}, we provide numerical
results obtained from several experiments done on Grid'5000 (we have
used up to 308 machines). We finish this paper with some prospective and
a conclusion, respectively in section~\ref{section:synthesis}
and~\ref{section:conclusion}.

\section{Distributed large scale systems}\label{section:grid}
\subsection{Grid Computing}

In~\cite{f02}, Foster and Kesselman define grid computing as
follows: ``A computational grid is a hardware and software
infrastructure that provides dependable, consistent, pervasive, and
inexpensive access to high-end computational capabilities''.

Thus, a Grid Computing or a Computational grid is a hardware and
software infrastructure allowing the sharing of a big number of
heterogeneous resources thanks to connection between several sites. The
resources sharing objective is to resolve problems confronted in
organisations which are often multi-sites and require an important
volume of data and an important computation power. These organisations
are called virtual organisations (VO). A computational grid is analogous
to the electric network which permits to any subscriber, at any time, to
accede instantaneously to the electric resource whatever is its origin
or location, via standardised interface~\cite{fkt01,fknt02}. But the
grid computing offers more services than the electric grid and should
guarantee criteria of reliability, security and access transparency
while taking account of constraints of the high throughput and the
choice of the quality of service (QoS).

\subsection{Peer-to-peer System}

One definition of a peer-to-peer system is: ``peer-to-peer refers to a
class of systems and applications that employ distributed resources to
perform a critical function in a decentralised manner''~\cite{ah01}.

Exchanges between systems can carry on the information, the processors
cycles, the memory or the files storage on disk. Contrary to the
client/server model, each node is a network entity which has the roles
of the server and the client at the same time. With peer-to-peer, the
personal computer can be part of the network. The peer-to-peer concerns
a class of applications which require hardware or human resources
available on the Internet. We distinguish two types of peer-to-peer
systems: 1) files sharing systems such as Gnutella, Napster and Kazaa,
which knew a great success on Internet and 2) intensive computation
oriented systems, equivalent to computational grids, such as
SETI@Home~\cite{url-seti}, XtremWeb~\cite{url-xtremweb} and
XtremWeb-CH~\cite{url-xtremwebch}.

\section{Desktop Grid}\label{section:desktop-grid}

Grids aim at providing a powerful infrastructure with quality-of-service
(QoS) guarantees to average size, homogeneous resources and certified
communities. In contrast, Peer-to-Peer systems focus on constructing a
very large infrastructure from larger communities of entrusted,
anonymous individuals and volatility resources. However, the convergence
of the two systems seems natural~\cite{tt03,fi03}. In fact, P2P research
focuses more and more on providing infrastructure and diversifying the
set of applications; Grid research is starting to pay attention to
increasing scalability. Desktop Grid combines the two concepts.

In this context, we aim at developing a new decentralised desktop grid
middleware using the features offered by several peer-to-peer tools as
Bonjour, Avahi and Free-Pastry. We will not build a new middleware from
scratch, but we would choose the most adequate protocol from these three
ones to build a decentralised middleware. Remark that others protocols
exist such as CAN~\cite{retc01} and CHORD~\cite{setc01}, but we choose
these three protocols because Bonjour and Avahi are two implementations
of Zero-configuration which already proved good success in local area
networks and for small organisations, whereas Free-Pastry, which is
very similar to CAN and CHORD, is based on DHT (Distributed Hash Table).

In next section, we expose how we can build a Desktop Grid by using
Peer-to-Peer technology.

\section{Using Peer-to-Peer techniques to build Desktop
  Grid}\label{section:peer-to-peer}

To provide a powerful Desktop Grid, it is important to have an important
number of resources. Therefore, it is necessary to integrate resources
made available by several institutions. The bottleneck, that limits the
scalability of such systems, is the centralisation character of existing
tools (see~\cite{url-xtremweb,url-xtremwebch} for the XtremWeb platforms
or~\cite{url-boinc} for the Boinc platform). Thus, it is primordial that
grids need more flexible distributed mechanisms allowing them to be
efficiently managed. Such characteristics are presented by Peer-to-Peer
systems, which proved their performance and ability to manage very big
number of interconnected peers in a decentralised manner. In addition,
theses systems support high volatility of resources.

Below, we describe three Peer-to-Peer systems, Bonjour, Avahi and
Free-Pastry, which are the candidates of our experiments tests.

\subsection{Bonjour}

Bonjour, also known as zero-configuration networking, enables automatic
discovery of computers, devices, and services on IP networks. Bonjour
uses industry standard IP protocols to allow devices to automatically
discover each other without the need to enter IP addresses or configure
DNS servers. Furthermore, Bonjour can allocate IP addresses without a
DHCP server, can translate between names and addresses without a DNS
server and can locate or advertise services without using a directory
server.

As a technical level, zero-configuration is a combination of three
technologies: link-local addressing, Multicast DNS, and DNS Service
Discovery. Link local addressing is viewed a safety net. When DHCP fails
or is not available, link-local addressing lets a computer make up an
address for itself, so that it can, at least, communicate on the local
link, even if wider communication is not possible. Like link-local
addressing, Multicast DNS is a safety net, so that when conventional DNS
servers are unavailable, unreachable, badly configured or otherwise broken,
computers and devices can still refer to each other by name in a way
that is not dependent on the correct operation of outside
infrastructure. DNS Service Discovery is built on top of DNS. It works
not only for with Multicast DNS (for discovering local services) but
also with good old-fashioned, wide-area Unicast DNS (for discovering
remote services).
 
\subsection{Avahi}

Avahi is a system which facilitates service discovery on a local
network. This means that you can plug your laptop or computer into a
network and instantly be able to view other people you can chat with,
find printers to print or find files being shared. Avahi is mainly based
on mDNS implementation for Linux. It allows programs to publish and
discover services and hosts running on a local network with no specific
configuration.

Avahi is an Implementation of DNS Service Discovery and Multicast DNS
specifications for Zero-configuration Networking. It uses D-Bus for
communication between user applications and a system daemon. The daemon
is used to coordinate application efforts in caching replies, necessary
to minimise the traffic imposed on networks.

\subsection{Free-Pastry}

Free-Pastry is a generic, scalable and efficient substrate for
peer-to-peer applications. Free-Pastry nodes form a decentralised,
self-organising and fault-tolerant overlay network within the
Internet. Free-Pastry provides efficient request routing, deterministic
object location and load balancing in an application-independent
manner. Furthermore, Free-Pastry provides mechanisms that support and
facilitate application-specific object replication, caching, and fault
recovery.

Free-Pastry performs application-level routing and object location in a
potentially very large overlay network of nodes connected via the
Internet. It can be used to support a variety of peer-to-peer
applications, including global data storage, data sharing, group
communication and naming.

Each node in the Free-Pastry network has a unique identifier
(nodeId). When presented with a message and a key, a Free-Pastry node
efficiently routes the message to the node with a nodeId that is
numerically closest to the key, among all currently live Free-Pastry
nodes. Each Free-Pastry node keeps track of its immediate neighbours in
the nodeId space, and notifies applications of new node arrivals, node
failures and recoveries. Free-Pastry takes into account network
locality; it seeks to minimise the distance messages travel, according
to a scalar proximity metric like the number of IP routing
hops. Free-Pastry is completely decentralised, scalable, and
self-organising; it automatically adapts to the arrival, departure and
failure of nodes.

\section{Description of the experimental
  setup}\label{section:experimental-setup}

Our goal is to study the scalability and the time response of the tools
described in the previous section. In fact, we focus on searching the
maximum number of supported registration nodes and the response time to
discover a given service. Note that the same benchmarks are applied for
the three Peer-to-Peer systems (Bonjour, Avahi and Free-Pastry). The
experimental platform is Grid'5000, highly reconfigurable and
controllable experimental grid platform gathering 9 sites geographically
distributed in France. Every site hosts a cluster from 256 CPU to 1K
CPU. All sites are connected by RENATER (10~Gb/s).

\subsection{Specific kernel on Grid'5000 }

Grid'5000~\cite{url-grid5000} offers an infrastructure with standard
kernels. To run our experimental test, it is necessary to customise one
kernel to support Avahi, Bonjour and Free-Pastry. Thus, we create a
specific kernel containing the entire needed package to run our codes
(registration and discovering codes for each system). After that, using
the two tools OAR and Kadeploy (see~\cite{url-oar,url-kadeploy}), we
reserve and we deploy this specific kernel in all the reserved
machines. We use only one site and all machines are made with AMD
Opteron processors with a 1~Gb/s network card.

\subsection{Sequential registrations}
	
In this test, the first step is to reserve N nodes on Grid'5000 (N will
vary from 100 nodes until a value for witch we observe a saturation of
the registration service). The number N represents the maximum nodes
that can be used for the experiment. Each node requests a registration
for a given service at given time. Initially, all nodes have the needed
codes to request a service but are inactive. Let $\delta$ be the
activation time. We activate sequentially all the requests (and we
receive back an acknowledgement). Indeed, the $k^{\mathrm{th}}$ request
will be activated at time $k\times\delta$. We increase $\delta$ to
analyse the behaviour of the system when the delay between events
becomes larger.

Obviously, at the beginning the number of registration is small, thus
the time of registration will be fast. We increase N until the
saturation value (i.e. the registration service no longer responds for a
new registration). We aim at analysing the scalability of the system
without overloading the network: in this test, only one multicast
appears at a given time.

\subsection{Simultaneous registrations}

In the first test, the registrations are done sequentially. This leads
to a limited number of communications to exchange information. In this
experiment, we stress the scalability of the system and its capacity to
manage the communications between the registered nodes. Therefore, we
request N (the number of reserved nodes) simultaneous registrations and
we compute the time to complete the registration step. If we obtain a
``reasonable'' response time, we increase N until the saturation
value. In others words, we are looking for the maximum registered nodes
that the system handle when the network is overloaded by several
multicast packet headers at the same time.

\subsection{Periodic registrations}\label{subsec:periodic-registration}
 
It's also important to study the efficiency of the system when there are
some nodes with the high volatility property. In such case, the system
needs to be updated by sending the global state to each node.

To simulate such behaviour, we register $N$ nodes then we cancelled
$\psi$ services ($N>\psi$) and we register them again randomly. It is
clear that the value of $\psi$ influences the efficiency of the
system. Therefore, we modify this value to obtain the maximum value for
the volatility of nodes.

\subsection{Real registrations}\label{subsec:real-registration}

In the periodic registration experiment, we simulate only one
disconnection/registration and this does not correspond to the real
behaviour of the operational grid systems since disconnections are more
frequent. In this test the same set of nodes is connected/disconnected
for several times. In this context, we are approaching the behaviour of
P2P systems and we measure the efficiency of such systems if they
interact as real grid system.
 
\subsection{Browsing services}

The other important metric is the time needed to browse a given
service. Indeed, in all the previous tests, we compute the registration
time. We need also to compute the discovering time which is the elapsed
time between the end of the registration of a unique service and the
date at which a browser node has discovered the service.

Note that the response time depends on the replicas number of a given
services and the registered nodes. The browsing program listen any new
event, i.e. a new registration or deleting services. With the four setup
mentioned before, we can analyse the performance of the discovery
service of the system. We have also the possibility to increase the
number of browsers. We draw the chart where point $(i,j,k)$ is the
response time $i$ for browser $j$ when we use a total of $k$ browsers.

\section{Performance of registration services
  analysis}\label{section:registration-analysis}

\subsection{Registration of Bonjour services}

\includefigurea{figure1}{Elapsed time for simultaneous registrations of
  Bonjour services}{simultaneous-registrations-bonjour}

\includefigureb{figure2}{Elapsed time for sequential registrations of
  Bonjour services}{sequential-registrations-bonjour}

Figure~\ref{fig:simultaneous-registrations-bonjour} corresponds to a
simultaneous registration. At the instant t we launch a registration of
one service on each machine (the activation time is the same for all
registrations). With up to 308 machines, the elapsed time of
registration varies between 1017 and 2307~ms. Nevertheless, better
registration times are given by the sequential test. In this test, every
minute, we activate a registration request of a
service. Figure~\ref{fig:sequential-registrations-bonjour} shows that
the elapsed time for the majority of services is between 1015 and
1030~ms. We mention that Bonjour does not show saturation until 308
machines in both simultaneous and sequential registration.

\subsection{Registration of Avahi services}

\includefigurea{figure3}{Elapsed time for simultaneous registrations of
  Avahi services}{simultaneous-registrations-avahi}

\includefigureb{figure4}{Elapsed time for sequential registrations of
  Avahi services}{sequential-registrations-avahi}

Figure~\ref{fig:simultaneous-registrations-avahi}
and~\ref{fig:sequential-registrations-avahi} point out that Avahi has
almost the same registration time in both simultaneous and sequential
tests. The elapsed time varies between 760 and 1110~ms. Comparing to
Bonjour, Avahi has better registration times. Note that with up to 308
machines Avahi has not been saturated.

\subsection{Registration of Free-Pastry services}

\includefigurea{figure5}{Elapsed time for simultaneous registrations of
  Free-Pastry services}{simultaneous-registrations-pastry}

\includefigureb{figure6}{Elapsed time for sequential registrations of
  Free-Pastry services}{sequential-registrations-pastry}

Contrarily to Avahi and Bonjour, Free-Pastry shows a big difference
between sequential and simultaneous registration tests. Indeed,
figure~\ref{fig:simultaneous-registrations-pastry} shows that in the
simultaneous registration, until the 160\up{th} service, the elapsed time
varies between 600 and 1000~ms. Beyond, the registration time increases
from one registration to another to reach 320000~ms.

On the other hand, as shown in
figure~\ref{fig:sequential-registrations-pastry}, the sequential test
demonstrates when activating just one request at each minute, we obtain
better time registration. Besides, the elapsed time for the registration
is almost stable and varies since the 60\up{th} service from 500 to
1000~ms. Like Bonjour and Avahi, the 295 registration (one registration
on one machine) doses not saturate Free-Pastry.

\section{Performance of discovery of services
  analysis}\label{section:discovery-analysis}

The second metric is to measure the necessary time to browse a
registered service. Then for each system (Bonjour, Avahi and
Free-Pastry) we measure the elapsed time between the termination of
registration and the discovery time. We repeat the same benchmarks for
both simultaneous and sequential registration. For that, we dedicate one
machine which runs the browser to discover services.

\subsection{Discovery behaviour of Bonjour}

Bonjour proves a good performance in discovering services. In fact, it
is able to discover 307 services registered on 307 machines (one service
on one machine). Furthermore, the discovery time doses not exceed 1
second. We mention that for the sack of simplicity and clearness we did
not put the numeric results in this paper for these two tests.

\subsection{Discovery behaviour of Avahi}

\includefigurea{figure7}{Browse services after a simultaneous
  registration}{simultaneous-browsing-avahi}

\includefigureb{figure8}{Browse services after a sequential
  registrations}{sequential-browsing-avahi}

Avahi loses almost 60\% of registered services. Besides, the discovery
time increases beyond 49 registrations to reach 900 s on
73\up{rd}. After that, the browse program of avahi spends about 220
seconds to find a registered service (see
figure~\ref{fig:simultaneous-browsing-avahi}).

Conversely, when we browse services after a sequential registration
(each minute, we request a registration), Avahi is able to discover more
services (303 from 307 registered services). Furthermore, the discovery
time is much better (less than 2 seconds for all services). Only for one
service did the discovery program spend 4203 seconds to find a service
(see figure~\ref{fig:sequential-browsing-avahi}).

\subsection{Discovery behaviour for Free-Pastry}

Free-Pastry has a good response time of services discovery but it loses
several services. In fact, we registered 306 services (one service on
one machine) and we launched the browsing program on another
machine. The browser can discover the service in at most one second. The
experimental tests show that, in sequential registration, the browser
finds just 270 from 293 services. Whereas, in simultaneous
registrations, the browser finds more services (275 from 292 services).

\section{Synthesis and future work}\label{section:synthesis}

\includefigurebig{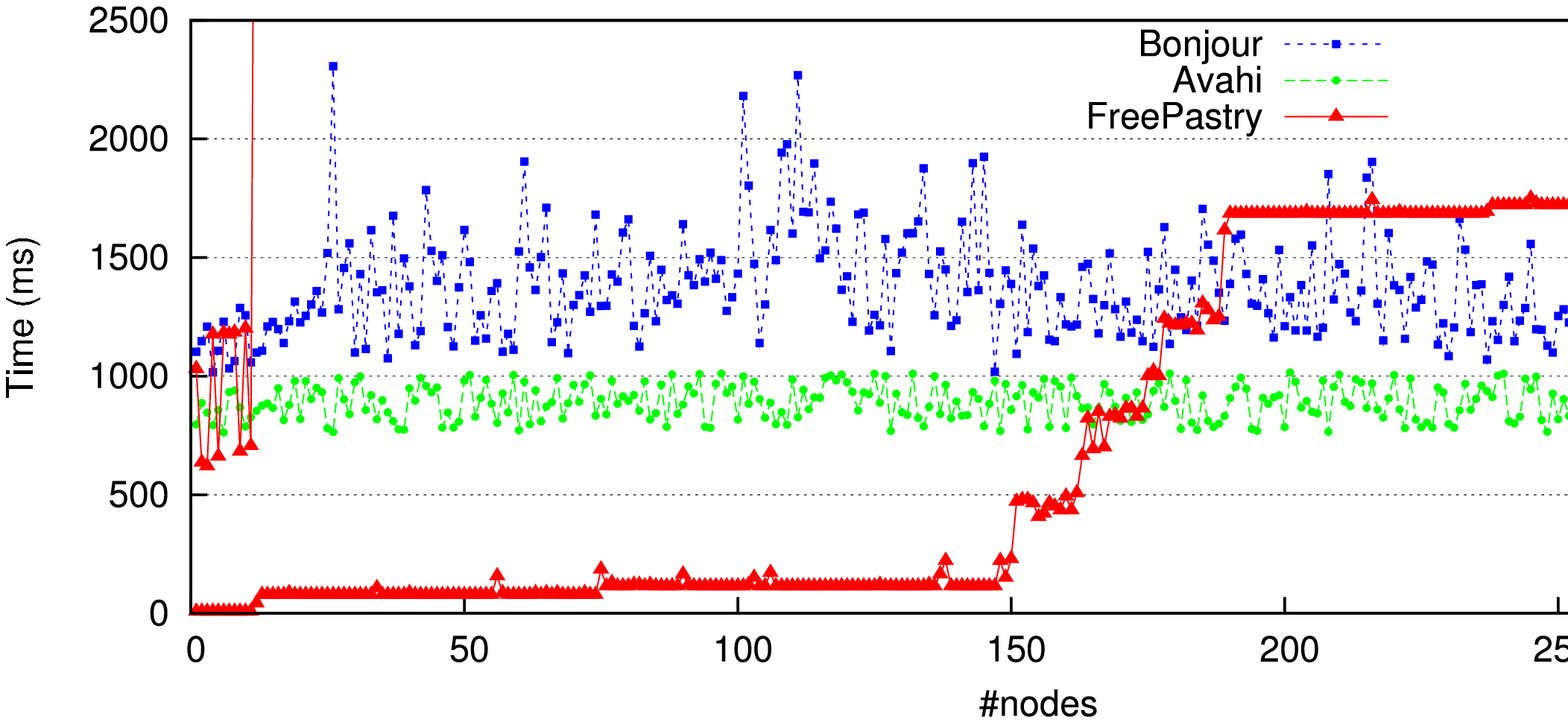}{Elapsed times for simultaneous registrations
  of services for Bonjour, Avahi and Free-Pastry (note that Free-Pastry is represented twice: once with the left-hand time axis, and a second time with the right-hand time axis, 140 times larger)}{combined-graph-simultaneous}

Figure~\ref{fig:combined-graph-simultaneous} illustrates the difference
between the three protocols: Bonjour, Avahi and Free-Pastry in the time
elapsed to register simultaneously about 300 services in a Peer-to-Peer
Network. Avahi is the best one because it spends the minimum times
(900--1000 ms). Bonjour needs more time to register services. However,
the difference is not important. Whereas Free-Pastry - which has times
near to Avahi up to 150 services -- needs distinctly higher times than
Avahi and Bonjour as shown in
figure~\ref{fig:combined-graph-simultaneous} (time corresponding to the
Free-Pastry curve are given on the secondary axe).

\includefigurebig{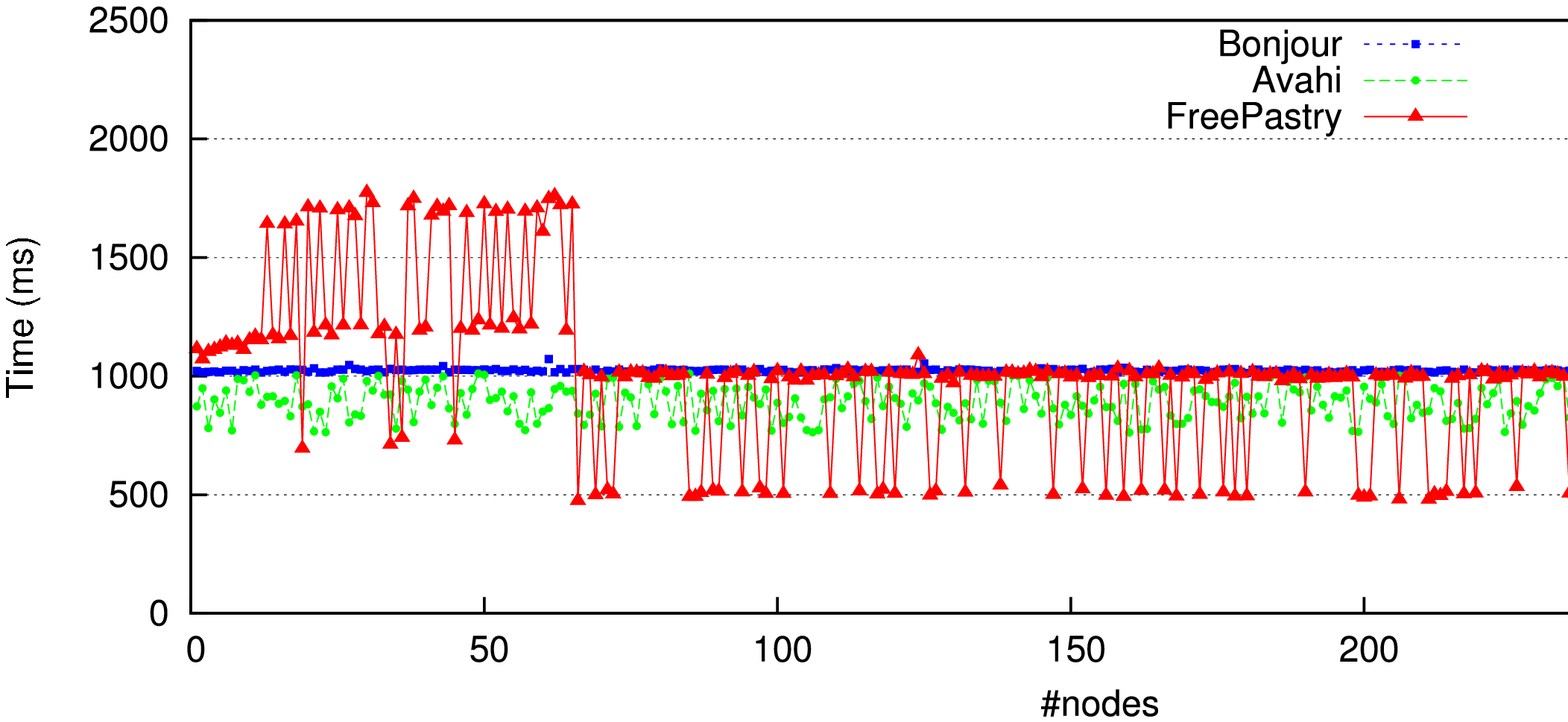}{Elapsed times for sequential registrations of
  services for Bonjour, Avahi and
  Free-Pastry}{combined-graph-sequential}

Figure~\ref{fig:combined-graph-sequential} shows the performance of
these tools (Bonjour, Avahi and Free-Pastry) when we register
sequentially one service on each machine (we attend about 300 machines).
We can mention that there is not a big difference between the three
protocols. In fact, Bonjour and Avahi give very similar results. Whereas
Free-Pastry, need less time to register some services (30\% of the total
machines), need the same time for others machines (60\%) and more time
than Avahi and Bonjour for the rest (10\%).

The lessons learnt from experiments are the following for our future
work that aims to replace the centralised registration service of nodes
for a desktop Grid platform such as XtremWeb by a decentralised service:
\begin{enumerate}
\item None of the three tools are superior for all the criteria ; it
  remains to understand (for instance) why Free-Pastry performance
  decreases drastically for simultaneous registration after 160
  registrations and why Avahi loses almost 60\% of registered services
  as soon as we browse more than 75 services which is not a
  lot. Clearly, Avahi does not scale at present time.
\item A lot of ``clock-interval'' artifacts can be seen in Free-Pastry
  and Avahi, especially in
  figures~\ref{fig:sequential-registrations-pastry}
  and~\ref{fig:simultaneous-browsing-avahi}. These artifacts are due to
  the event model used in these protocols, where a ``watch loop'' does
  periodic polling of resources to be aware of state changes (and thence
  create notifications).
\item The Bonjour implementation that we have used seems to offer the
  best compromise. However the Bonjour API is not as rich as
  Free-Pastry's API. In Free-Pastry we have the PAST module that
  implements archival storage that could be used, in our context, for
  storing the properties (CPU Mhz, RAM available...) or data produce on
  nodes asking for participating to the desktop grid.
\end{enumerate}

Experiments and analyses of P2P networks have been conducted over the
Grid'5000 platform for the generic \jxta P2P framework~\cite{acdj07}.
In this article, the goal of the performed benchmarks is similar to our
goal. It concerns to answer common and unanswered questions: how many
rendezvous peers are supported by \jxta in a given group and what is the
expected time to discover resources in such groups?

Two main protocols of \jxta have been evaluated in~\cite{acdj07}: 1) the
peerview protocol used to organize super peers, known as rendezvous
peers, in a \jxta overlay and 2) the discovery protocol, that relies on
the peerview protocol, used to find resources inside a \jxta
network. All sites of Grid'5000 were used and a mix of hundreds of
rendezvous peers and normal peers, called edge peers, have been deployed
on at most 580 nodes.  Results show that with default values for
parameters of the peerview protocol, the goal of the algorithm is not
achieved, even with as few as 45~rendezvous peers.  However, parameter
tuning makes it possible to reach larger configurations in terms of
number of rendezvous peers. For the discovery protocol, authors show
that discovery time is rather smaller, provided that all rendezvous
peers satisfy a given property. These results give developers a better
view of the scalability of \jxta protocols.

Our results, augmented with those of~\cite{acdj07} clearly demonstrate
that for open source projects as well as for industrial software with
production quality, there is a strong need to test and evaluate the
properties of the distributed system in real scale platform such as
Grid'5000.

\section{Conclusion}\label{section:conclusion}

To launch computing work on the nodes of the grid, it is essential to
browse the nodes and then execute works. Furthermore, in the mean time,
it is possible that new nodes comes and want to register to the
grid. Then it is important to register this new machine in the minimum
time possible to be added to the grid.

In this paper, we have illustrated the performance of three P2P systems
which are Bonjour, Avahi and Free-Pastry. Several experimental tests
were executed on Grid'5000 where the used machines number could go until
308. Bonjour proves a high performance in registration services and is
able to browse all registered services (307 services) and, especially,
can discover them in at most 1 second each one. Avahi is powerful in
registration of new services but it is not able to browse all the
services (it loses 60\% services in simultaneous registration) and need
a long time (4200 seconds) to discover services which have been
registered in a sequential manner. On the other hand, Free-Pastry shows
good results in the sequential registration of 295 services (at most 1
second needed to register one service on each machine), but it spends
more time (300 seconds) to find a services when we activate 295
registration requests of new services at the same time.

Finally, we did not include the results for periodic and real
registrations (presented in the above
sections~\ref{subsec:periodic-registration}
and~\ref{subsec:real-registration}) because there is no important
difference in discovering services or registrations to be mentioned. In
fact, as illustrated before, the behaviour of discovering or
registration depends only of the registrations mode (sequential or
simultaneous).

\section*{Acknowledgement}

Experiments presented in this paper were carried out using the Grid'5000
experimental testbed, an initiative from the French Ministry of Research
through the ACI GRID incentive action, INRIA, CNRS and RENATER and other
contributing partners (see~\cite{url-grid5000}). We would like to thank
Mathieu Jan for his fruitful comment on an early version of this paper.

\bibliography{biblio}
\end{document}